\documentclass[epj,referee]{svjour}
\usepackage{pstricks}
\usepackage{amsmath}
\usepackage{graphicx}
\usepackage{graphics}
\usepackage{color}

\newcommand{\real}[1]{\Re \text{e}\left\{#1 \right\}}

\newcommand{\be}[1]{\begin{eqnarray}
#1
\end{eqnarray}}

\begin{document}

\title{Spectral properties and stability in the Two-Dimensional Lattice-Hubbard model.}
\author{C. A. Lamas\inst{1} }

\institute{Departamento de F\'{\i}sica, Universidad Nacional de La Plata,
C.C. 67, 1900 La Plata, Argentina}
\date{Received: date / Revised version: date}
%
\abstract{ The  two-dimensional Hubbard model on the square lattice
is studied in the presence of lattice distortions in the adiabatic
approximation. The self energy is computed within perturbation
theory up to second order, which provides a way for studying the
quasiparticle dispersion. We compute numerically the second order
contribution to the self-energy using a standard Fast Fourier
Transform Algorithm for finite sizes system. The stability of the
lattice distortions is investigated and a schematic phase diagram is
drawn. The Fermi surface is analyzed for densities close to half
filling, the presence of lattice distortions changes some spectral
properties of the model and gives an anisotropic interacting Fermi
surface. The spectral function is calculated along several lines in
momentum space and the renormalized quasiparticle dispersion is
obtained. The behavior of the density of states is shown for
different values of the intrasite repulsion $U$ in the different
phases.
} 
%


\maketitle

\section{Introduction}
The two dimensional Hubbard model  has been usually associated with
magnetism and superconductivity and is a promising toy model for the
electronic degrees of freedom of high-temperature superconductors.
The most interesting properties occur mostly in underdoped samples,
with electron densities close to half filling, where the system is
an antiferromagnetic Mott-insulator. The competition between the
kinetic and Coulomb terms gives rise to strong electron-electron
correlations.

The Peierls instability towards a spatially broken symmetry state in
one dimensional systems is caused by the competition between the
energy of lattice distortions and the formation of  gap at the Fermi
level in the electronic spectrum. This instability can occur in two
dimensions if the structure of the Fermi surface has a strong
nesting by a single vector. It is the case of the two-dimensional
square lattice in the tight-binding approximation at half filling.

Elastic Umklapp scattering with momentum transfer of
$(\pi,\pi)$\cite{nota1} across the Fermi Surface is allowed when it
extends to the Brillouin zone boundary at half filling. In this
case, a small electron-lattice coupling will induce a lattice
dimerization which is related with a periodic modulation of the bond
hopping called in the literature, bond-order wave (BOW). This
motivates the study of the typical patterns for elastic deformations
with modes corresponding to the nesting vector $(\pi,\pi)$.


Although a finite frequency for the phonons would be important to
study their influence in the mechanism of high Tc superconductivity,
considering them as adiabatic could shed light on their influence on
the different inhomogeneous phases that have been observed, e.g. in
the underdoped region. Besides, their role is expected to be
important for the undoped parent compounds.

Electron phonon coupling can lead to charge inhomogeneities, such as
stripes, as has been studied in \cite{moreo} within a spin-fermion
model. However, this happens far from half-filling and for
sufficiently strong diagonal coupling.

On the other hand, the lattice distortions change the Fermi Surface
(FS) shape and the FS deformation due to the presence of
interactions is a central question within the breakdown of the Fermi
liquid theory. \cite{luttinger1,luttinger2,luttinger3,lamas}

In the strong coupling regime one can find not only a deformed FS,
but may even find a different topology. The FS of the non
interacting Hubbard model is closed around the origin in the
reciprocal space while in the interacting case it can be a surface
closed around the point $(\pi,\pi)$. It is in this regime where the
role of the elastic distortions can take an important place in the
high temperature superconductor phases \cite{anderson}

In this paper we study the stability of lattice distortions in the
presence of Coulomb interactions $U$. The critical values $U_{c}$
where the distortions are suppressed are studied and a schematic
phase diagram is drawn. The spectral properties  are analized in the
region of parameter space where the elastic deformations  are
favored and the FS shape is constructed from the renormalized
dispersion. The FS shape was studied for the Hubbard model in the
last years
\cite{valenzuela,nojiri,shonhammer,metzner-rohe,metzner,metzner_letter},
but the influence of lattice distortions has not been taken into
account. The Polaron formation in the Holstein-Hubbard model was
investigated in \cite{fehske} by means the slave-boson saddle-point
approximation and the $t-J$ model with electron-phonon interaction
was studied recently \cite{riera} for small systems. We show that
these distortions have an influence in the FS shape leading to an
anisotropic FS in the $\delta\neq 0$ phase.

\section{Second order Perturbation theory.}

\subsection{The model}

We examine the Hubbard model, coupled  with a classical phonon field
, which describes spin-$\frac{1}{2}$ fermions in a two dimensional
square lattice with nearest-neighbor interactions, in the presence
of lattice distortions in the adiabatic approximation. This model
represent a useful toy model to describe the physic in the presence
of vibrational modes . The dependence of the hopping amplitudes
$t_{i,j}$ is assumed to be linear in the lattice distortions.
 \begin{eqnarray}\label{hubbard}
    \nonumber H&=&-t_{0}\sum_{\sigma}\sum_{i=1}^{L}\sum_{j=1}^{L}
    (1+\alpha(u^{x}_{i+1,j}-u^{x}_{i,j}))\;c_{i,j,\sigma}^{\dagger}c_{i+1,j,\sigma}\\
    &+&(1+\alpha(u^{y}_{i,j+1}-u^{y}_{i,j}))c_{i,j,\sigma}^{\dagger}c_{i,j+1,\sigma}+\text{H.c})\\
    \nonumber
    &+&  U\sum_{i,j}n_{i,j\uparrow}\;n_{i,j\downarrow}\\ \nonumber
    &+&\frac{K}{2}\sum_{i,j}((u^{x}_{i+1,j}-u^{x}_{i,j})^{2}+(u^{y}_{i,j+1}-u^{y}_{i,j})^{2}),
 \end{eqnarray}
\begin{figure}[ht]
  \includegraphics[width=0.45\textwidth]{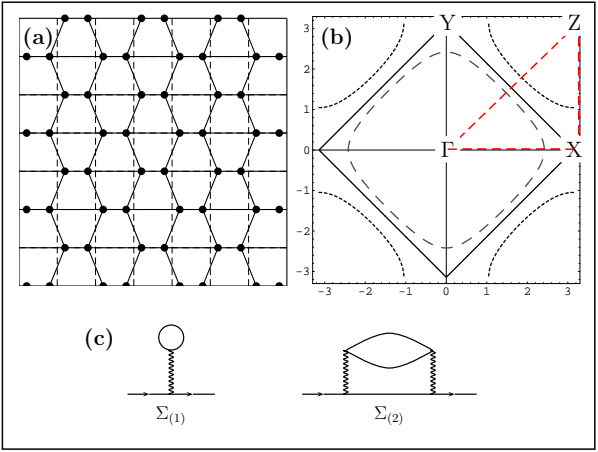}\\
  \caption{\label{fig:varios}\textbf{a)} Lattice shape for the modulations corresponding to
  pattern (a). \textbf{b)} Fermi surfaces for various densities $\rho$ in the two-dimensional
   free fermion case. For densities $\rho<1$ the FS is centered in the  $\Gamma$-point and
    for $\rho>1$ it is centered in the $Z$-point.
    \textbf{c)} Diagrams which contribute to the self energy up to second order in $U$. The wavy lines
  represent the on site repulsive interaction and the solid lines the free propagators.}
\end{figure}
where $t_{0}$ is the transfer integral between nearest neighbor
sites in the absence of distortions,
$n_{i,j,\sigma}=c^{\dagger}_{i,j,\sigma}c_{i,j,\sigma}$ is the
number operator for electrons of spin $\sigma$ at site $i$, $\alpha$
is the electron-phonon coupling, $K$ is the elastic constant of the
lattice and $U$ is the on site Hubbard interaction. It is convenient
to work with the dimensionless parameters
\begin{eqnarray}
\delta^{x}_{i,j}&=&\alpha(u^{x}_{i+1,j}-u^{x}_{i,j})\\
\delta^{y}_{i,j}&=&\alpha(u^{y}_{i,j+1}-u^{y}_{i,j})\\
\lambda &=& \frac{\alpha^{2}t_{0}}{K}
\end{eqnarray}
and in the following we fix the energy scale setting $t_{0}=1$.

We first review briefly the free case $(U=0)$ and
$\delta^{x,y}\equiv 0$, where:
\begin{eqnarray}\label{freeband}
    \varepsilon_{0}(\vec{k})=-2(\cos k_{x}+\cos k_{y}).
 \end{eqnarray}
The FS for different densities $\rho$ is shown in Figure
\ref{fig:varios}-b, for half-filling $(\rho=1)$ being a diamond
centered at the origin with vertices at $\pm(0,\pi)$ and
$\pm(\pi,0)$.

Let us consider the two possible alternation patterns consistent
with the nesting vector $Q=(\pi,\pi)$:
 \begin{eqnarray}\label{pattern a}
    \nonumber \delta_{i,j}^{x}=(-1)^{i+j}\delta_{0}^{x}\hspace{1cm}\text{or}\hspace{1cm} \delta_{i,j}^{x}=(-1)^{i+j}\delta_{0}^{x}\\
    \delta_{i,j}^{y}=0 \hspace{1cm}\;\;\hspace{1cm}  \delta_{i,j}^{y}=(-1)^{i+j}\delta_{0}^{y},
 \end{eqnarray}
namely patterns (a) and (b) respectively \cite{nota2}. The two
dimensional Peierls instability in the pure-hopping case was studied
by Tang and Hirsch  \cite{hirsch}  founding that at half filling
pattern (a) is favored. Using a Mean Field (MF) approach we can see
that for a wide range of values of $U$ this pattern survives up to a
critical value $U_{c}$. First we review the MF results and later, by
using second order perturbation theory we reexamine the stability of
the distortions in the present model.

When one considers a non trivial Hubbard local interaction, it is
easy to see, using Hartree-Fock \cite{yuan}, that for $U<U_{c}$
distortions (a) are favored at half filling and for $U>U_{c}$ the
distortions are quickly suppressed.

Treating the  local Coulomb repulsion in the MF approximation,
\begin{eqnarray}
   \nonumber U\sum_{i,j}n_{i,j,\uparrow}n_{i,j,\downarrow}\rightarrow
    &U&\sum_{i,j}( \langle n_{i,j,\uparrow}\rangle n_{i,j,\downarrow}+
    \langle n_{i,j,\downarrow}\rangle n_{i,j,\uparrow}\\
    &-&
    \langle n_{i,j,\uparrow}\rangle \langle
    n_{i,j,\downarrow}\rangle)
\end{eqnarray}
the hamiltonian becomes quadratic and it is straightforwardly
diagonalized in reciprocal space.

The expectation value $\langle n_{i,j}\rangle $ is assumed uniform,
and the electron density with a given spin can be assumed as
$\langle n_{i,j,\sigma}\rangle = \frac{1}{2}+\frac{\sigma}{2}
(-1)^{i+j}m$ \cite{yuan}, $m$ being the staggered magnetization.
Within this approximations the electronic spectra for the two
patterns (a) and (b) are
\tiny
\begin{eqnarray}
   \varepsilon^{(a)}(\textbf{k})_\pm &=&\pm
2t\sqrt{\frac{U^{2}m^{2}}{4}+4(\cos k_{x} + \cos k_{y}
)^{2}+4\delta^{2}\sin^{2} k_x }\\
\nonumber \varepsilon^{(b)}(\textbf{k})_\pm &=&\pm
2t\sqrt{\frac{U^{2}m^{2}}{4}+4(\cos k_{x}+\cos
k_{y})^{2}+4\delta^{2}(\sin k_x + \sin k_y )^{2}}
\end{eqnarray}
\normalsize

The ground state energies for the two patterns are
\begin{eqnarray}
    E=\frac{2}{L^{2}}\sum_{k}\varepsilon^{a(b)}(k)_{-}+\frac{U}{4}(1+m^{2})+r\frac{\delta^{2}}{2\lambda}
\end{eqnarray}
with $r=1$ for pattern (a) and $r=2$ for pattern (b).

Minimizing the energy with respect to $\delta$ and $m$ for various
values of $U$ and $\lambda$ we can easily see that pattern (a) has
lower energy than pattern (b) and for a huge range of $U$ the values
of $\delta^{*}_{MF}$ for minimum energy are almost constant.

\begin{figure}[t]
  \includegraphics[width=0.48\textwidth]{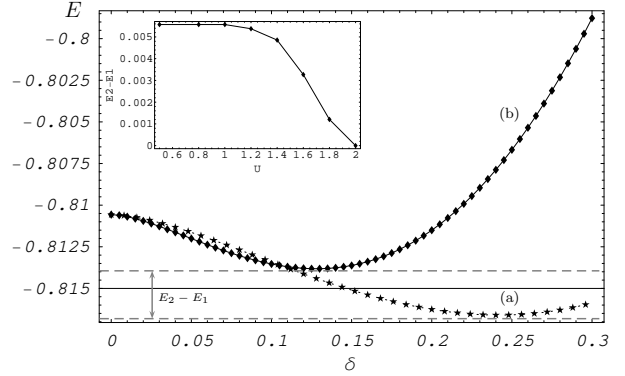}\\
  \caption{\label{fig:ayb} Energy calculated for both patterns (a) stars and (b) squares
   as a function of $\delta$ for $U=0.4$, $\lambda=1/2$ and $m=0$. Clearly pattern (a)
   has lower energy. On the other hand it is easy to see that the value of $\delta$ in the
   minimum energy is almost constant in the range of $U$ between $0$ and $\sim2$. In the Inset the
   difference of energy between the two patterns is shown as a function of $U$, this difference
   falls off to zero at $U_{c}\approx 2$ where the distortion $\delta$ its suppressed.      }
\end{figure}

For example for $\lambda=0.5$ we have for pattern (a)
$\delta^{*}_{MF}\simeq0.24$ and for pattern (b)
$\delta^{*}_{MF}\simeq0.13$. The value of the interaction where
the distortions are suppressed is $U_{c}\sim 2$.

In the presence of distortions that follow pattern (a) the lattice
changes as we show in Figure \ref{fig:varios}-a. The unit cell is
doubled and the Brillouin Zone is reduced to half. It is
straightforward to find the electronic spectra for the deformed free
case \ \ ($U=0$). Working in the first Brillouin zone (BZ) we have
two single particle bands
 $$\tilde{\varepsilon}_{0}(\textbf{k})_\pm =\pm
2t\sqrt{(\cos k_{x} + \cos k_{y} )^{2}+\delta^{2}\sin^{2} k_x}.$$
for pattern (a) and
 \small
 $$\tilde{\varepsilon}_{0}(\textbf{k})_\pm =\pm
2t\sqrt{(\cos k_{x} + \cos k_{y})^{2}+\delta^{2}( \sin^{2} k_x +
\sin^{2} k_y )}.$$
 \normalsize
for pattern (b). In the following we will work in the extended
Brillouin zone formed by a square centered in the point
$\Gamma=(0,0)$ with vertices in $\pm Z$ and $\pm \bar{Z}$ where
$Z=(\pi,\pi)$ and $\bar{Z}=(\pi,-\pi)$. The unperturbed dispersion
in the extended BZ is
\scriptsize
\begin{eqnarray}
\varepsilon(\textbf{k})&=&2(1-2\;\theta(\eta(\textbf{k})))\sqrt{\eta(\textbf{k})^{2}+\delta^{2}\sin^{2}
k_{x}}\\
\varepsilon(\textbf{k})&=&2(1-2\;\theta(\eta(\textbf{k})))\sqrt{\eta(\textbf{k})^{2}+\delta^{2}(\sin^{2}
k_{x}+\sin^{2} k_{y})}
\end{eqnarray}
\normalsize
for patterns (a) and (b) respectively, where we have used
{$\eta(\textbf{k})=\cos k_{x}+\cos k_{y}$} and $\theta$ is the
step function. In order to fix the notation we present here other
important points used in this work like $X=(\pi,0)$, $Y=(0,\pi)$
and $M=(\frac{\pi}{2},\frac{\pi}{2})$.

\subsection{Second order self energy}

The  contributions to the self energy are given by the diagrams
shown in Figure.\ \ref{fig:varios}-c, these contributions were
calculated using ordinary perturbation theory and were calculated
slightly away half filling where divergences can appear. To deviate
slightly half-filling are not significant changes in patterns of
deformations in the mean field treatments and previous work showing
that there is no appearance of stripes phase for fillings close to
one \cite{moreo}.

 Since the partners in the scattering processes have opposite spins, other
possible diagrams with two Coulomb lines are absent.

The contributions of these diagrams are given by
 \be{%
 \nonumber \Sigma^{(1)}(\textbf{k},\omega)&=&-U\frac{\imath}{(2\pi)^{3}}\int_{-\infty}^{\infty}
 d\omega'\int_{BZ}d^{2}\textbf{k}'G^{(0)}(\textbf{k}',\omega')\\
 &=&\frac{U\;\rho}{2} \ ,
 }%
where $\rho$ is the electronic density of the system and
%
%
\footnotesize
 \be{
 \Sigma^{(2)}(\textbf{k},\omega)&=&\frac{(U)^{2}}{(2\pi)^{6}}
 \int_{-\infty}^{\infty} d\omega' \int_{-\infty}^{\infty} d\omega''
 \int_{BZ}d^{2}\textbf{k}'\int_{BZ}d^{2}\textbf{k}''\times\\
 \nonumber&\times& G^{(0)}(\textbf{k}',\omega') G^{(0)}(\textbf{k}'',\omega'')
  G^{(0)}(\textbf{k}+\textbf{k}'-\textbf{k}'',\omega+\omega'-\omega'') \ ,
}
 \normalsize
where
 \be{%
 G^{(0)}(\textbf{k},\omega)=\frac{1}{\omega-\xi_{\textbf{k}}+\imath\gamma_{\textbf{k}}}
  }%
with $\xi_{\textbf{k}}=\varepsilon_{0}(\textbf{k})-\mu$,
$\gamma_{\textbf{k}}=\gamma \; {\rm sign} (\xi_{\textbf{k}})$ and
$\mu$ is the chemical potential.

To first order, the contribution of $\Sigma^{(1)}$ is a real
constant  ($\textbf{k}$-independent) that shifts the dispersion
relation and does not contribute to the FS deformation because it
can be absorbed by a shift in the chemical potential $\delta
\mu_{1}=Un/2$ to keep the density fixed.
\begin{figure}[t]
  \includegraphics[width=0.48\textwidth]{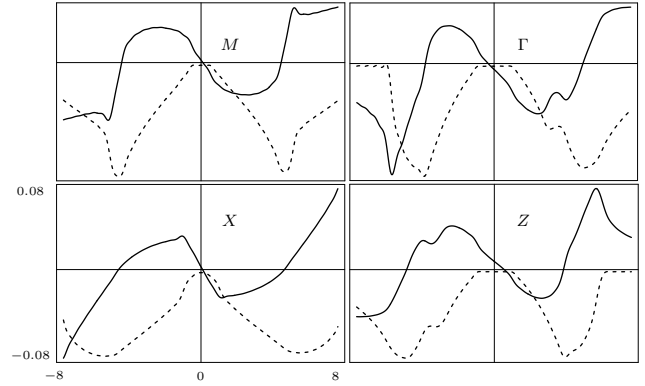}\\
  \caption{\label{fig:sigmavsomega} Self Energy in $M$, $\Gamma$,$X$ and $Z$ points at $\rho=0.995$,
   The solid line correspond to the real part and the dotted one to the imaginary part of the Self Energy.    }
\end{figure}
The second order contribution to the self energy
%
 \be{%
 \nonumber \Sigma^{2}(\textbf{k},\omega)&=&\frac{U^{2}}{L^{2}}
 \sum_{\textbf{k}',\textbf{k}''}
\;
 \left(\frac{\theta(\xi_{\textbf{k}-\textbf{k'}})   \theta(\xi_{\textbf{k'}-\textbf{k''}})
  \theta(-\xi_{\textbf{k''}})}
 { \omega -
 \xi_{\textbf{k}-\textbf{k'}}-\xi_{\textbf{k'}-\textbf{k''}}+\xi_{\textbf{k''}}}\right.\\
 &+&\left.
\frac{\theta(-\xi_{\textbf{k}-\textbf{k'}})
\theta(-\xi_{\textbf{k'}-\textbf{k''}}) \theta(\xi_{\textbf{k''}})}
 { \omega - \xi_{\textbf{k}-\textbf{k'}}-\xi_{\textbf{k'}-\textbf{k''}}+\xi_{\textbf{k''}}} \right)
 }
%
 is $\textbf{k}$-dependent producing a renormalized dispersion and
leads to a FS deformation. This term is computed numerically in what
follows using a Fast Fourier Transform (FFT) for finite size
systems. The momenta in the Brillouin zone are discrete and defined
by $k_{x,y}=-\pi+\Delta k (n_{x,y}-1)$ and $\Delta
k=\frac{2\pi}{L}$, $n_{x,y}=1,\ 2,\ \dots ,\ L$

\begin{figure}[t]
  \includegraphics[width=0.48\textwidth]{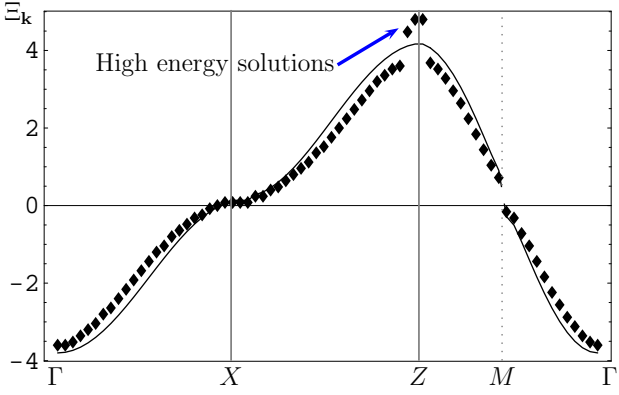}\\
  \caption{\label{fig:renor} Renormalized dispersion  relation corresponding to
  $\delta=0.24$ and $U=1.8$
  in the $\Gamma\rightarrow X\rightarrow Z\rightarrow \Gamma$ line.
   The squares correspond to the renormalized dispersion calculated
   by means of Eq.\ (\ref{eq:secular}) and the solid line corresponds to the free case.
    We can see that the Coulomb interaction reduces the bandwidth. In the $Z$-point some high energy
    solutions appear.  }
\end{figure}

The retarded self energy in space-time representation is
 \begin{eqnarray}\label{retarded}
    \Sigma^{(2)}_{ret}(\textbf{x},t)&=&\mathcal{F}_{\textbf{k}}\mathcal{F}_{\omega}
    \left[\Sigma^{(2)}(\textbf{k},\omega+i 0^{+})\right]\\ \nonumber
    &=&\frac{1}{L}\sum_{\textbf{k}}\int_{-\infty}^{\infty}\frac{d\omega}{2\pi}
    \;e^{i(\textbf{k}\cdot\textbf{x}-\omega t)}\Sigma^{(2)}(\textbf{k},\omega+i 0^{+})
 \end{eqnarray}

where the Fourier transform is defined by
 \begin{eqnarray}\label{retarded}
   \mathcal{F}_{\textbf{k}}\left[g(\textbf{k})\right](\textbf{x})&=&
   \frac{1}{L}\sum_{\textbf{k}}e^{i\textbf{k}\cdot\textbf{x}}g(\textbf{k})\\
   \mathcal{F}_{\omega}\left[g(\omega)\right](t)&=&
    \int_{-\infty}^{\infty}\frac{d\omega}{2\pi}e^{-i\omega t}g(\omega)
 \end{eqnarray}
and the inverse transformation is denoted by
$\mathcal{F}_{\textbf{k}}^{-1}$ and $\mathcal{F}_{\omega}^{-1}$
respectively
The frequency Fourier transform is straightforward
\scriptsize
 \begin{eqnarray}\label{retarded}\nonumber
    \Sigma^{(2)}_{ret}(\textbf{x},t)&=&-\;i\;\theta(t)\frac{U^{2}}{L^{2}}
    \;\mathcal{F}_{\textbf{k}}\left[\sum_{\textbf{k}',\textbf{k}''}
    z(\textbf{k}-\textbf{k'},t)z(\textbf{k'}-\textbf{k''},t)w^{*}(\textbf{k''},t)\right.\\
    &+&\left.
    w(\textbf{k}-\textbf{k'},t)w(\textbf{k'}-\textbf{k''},t)z^{*}(\textbf{k''},t)\right]
 \end{eqnarray}
%
%
%
\normalsize
 where we have used the short-hand notation
$$\displaystyle{z(\textbf{k},t)=e^{-i\xi_{ \textbf{k}
}t}\theta(-\xi_{\textbf{k}})}$$
$$w(\textbf{k},t)=e^{-i\xi_{\textbf{k}}t}\theta(\xi_{\textbf{k}}).$$
The transformation in $\textbf{k}$ is the Fourier transform of a
convolution and can be written as product of the Fourier transforms
\small
 \begin{eqnarray}\nonumber
    \Sigma^{(2)}_{ret}(\textbf{x},t)=-\;i\;\theta(t)\frac{U^{2}}{L^{2}}
    \;\left[
    \tilde{z}^{2}(\textbf{x},t)\tilde{w}^{*}(\textbf{x},t)+
    \tilde{w}^{2}(\textbf{x},t)\tilde{z}^{*}(\textbf{x},t)\right]
 \end{eqnarray}
\normalsize
 with
$\tilde{z}(\textbf{x},t)=\mathcal{F}_{\textbf{k}}[z(\textbf{k},t)]$
and
$\tilde{w}(\textbf{x},t)=\mathcal{F}_{\textbf{k}}[w(\textbf{k},t)]$

Clearly $\tilde{z}(\textbf{x},t)=\langle c_{\textbf{x}}(t)
c^{\dagger}_{\textbf{0}}(0)\rangle$ and
$\tilde{w}(\textbf{x},t)=\langle
c^{\dagger}_{\textbf{0}}(0)c_{\textbf{x}}(t)\rangle$ and the
causal free propagator in space-time is given by
$G^{0}(\textbf{x},t)=-i(\theta(t)\tilde{z}(\textbf{x},t) -
\theta(-t)\tilde{w}(\textbf{x},t) )$

The functions $\tilde{z}(\textbf{x},t)$ and
$\tilde{w}(\textbf{x},t)$ can be calculated for finite size systems
with a standard FFT algorithm. Then, replacing the result in the
equation for $\Sigma^{(2)}_{ret}(\textbf{x},t)$, we can obtain the
self energy in momentum space by means of the inverse Fourier
transform
$$\Sigma^{(2)}(\textbf{k},\omega+i0^{+})=\mathcal{F}_{\textbf{k}}^{-1}
\mathcal{F}_{\omega}^{-1}[\Sigma^{(2)}_{ret}(\textbf{x},t)].$$

The real and imaginary parts of the self energy are shown in Figure
\ref{fig:sigmavsomega} for $M$, $\Gamma$, $X$ and $Z$ points at
$\rho=0.995$.
The shape of the Self Energy is very similar to the one found for
the Hubbard model \cite{zlatic} but some features are
$\delta$-dependent. For example, in Figure \ref{fig:sigmavsomega}
the real part of $\Sigma$ in the point $X$ has a clear linear
behavior in a wide interval around $\omega=0$. The range of $\omega$
where this occurs is larger for higher values of $\delta$. The
symmetries of the Hubbard model are preserved. We can see in Figure
\ref{fig:sigmavsomega} that if one changes $\omega\rightarrow
-\omega$ in the plot for the imaginary part of $\Sigma$ at point
$\Gamma$,  the one corresponding to point $Z$ is recovered. In the
plots for the points $\Gamma$ and $Z$ it can be seen that
$\real{\Sigma(\textbf{k},\omega)}=-\real{\Sigma(\textbf{k}+\textbf{Q},-\omega)}$,
with $\textbf{Q}=(\pi,\pi)$.

The interacting Green function \cite{mahan} can be calculated up to
second order

\begin{eqnarray}
G(\textbf{k},\omega+i0^{+})=\frac{1}{\omega+i0^{+}-\xi_{\textbf{k}}-\Sigma(\textbf{k},\omega+i0^{+})}
\end{eqnarray}
and the low energy excitations can be determined from the equation

\begin{eqnarray}\label{eq:secular}
G^{-1}(\Omega,\textbf{k})=\Omega-\xi_{\textbf{k}}-\Re e
\Sigma(\textbf{k},\Omega)=0.
\end{eqnarray}

For each point in the Brillouin zone, the last equation gives the
renormalized dispersion as a function of the chemical potential
\begin{eqnarray}
\Xi_{\textbf{k}}=\Omega.
\end{eqnarray}

In Figure \ref{fig:renor} we show the renormalized dispersion for
$U=1.8$ in the path shown in Figure \ref{fig:varios}. We can see
that the Coulomb interaction reduces the bandwidth.

For values of the momentum away from the FS, in particular for high
values of $U$, Eq.\ (\ref{eq:secular}) can give more solutions
corresponding to higher energy excitations. This is particularly
clear in the points $\Gamma$ and $Z$ where two new bands appear for
$U>1$. This is reminiscent of the Hubbard bands. Figure
\ref{fig:renor} shows the high energy solutions in the $Z$-point. We
chose to show a result for a high value of $U$ because the effects
of renormalization are more evident and then the high energy
solutions are visible.

\subsection{Stability of lattice distortions.}

The Ground State energy per site was calculated for several values
of $U$ and $\lambda$
\begin{eqnarray}\label{eq:GSenergy}
E=\frac{2}{(2\pi)^{2}} \int d^{2}\textbf{k}\;
\Xi_{\textbf{k}}(\delta)\theta(-\Xi_{\textbf{k}}(\delta))+r\frac{\delta^{2}}{2\lambda}
\end{eqnarray}
where $r=1$ for pattern $(a)$ and $r=2$ for pattern $(b)$

Keeping fixed the filling, we can calculate the energy for various
values of $\delta$ and $\lambda$ as a function of $U$. In this way
we can study the stability of the lattice distortions when
increasing the Coulomb interaction $U$, beyond the Mean Field
approximation. In the following we restricted ourselves to study
pattern (a) . The values of $\delta$ that minimize the total energy
depend of $\lambda$ and when $\lambda\rightarrow 0$, the distortions
are suppressed. In our calculations the deformations are suppressed
for values below some $\lambda$ finite since the calculations were
made for finite size systems (in general for systems of $120\times
120$ sites, but the spectral properties are not severely changed for
bigger sizes. )
\begin{figure}[t]
  \includegraphics[width=0.48\textwidth]{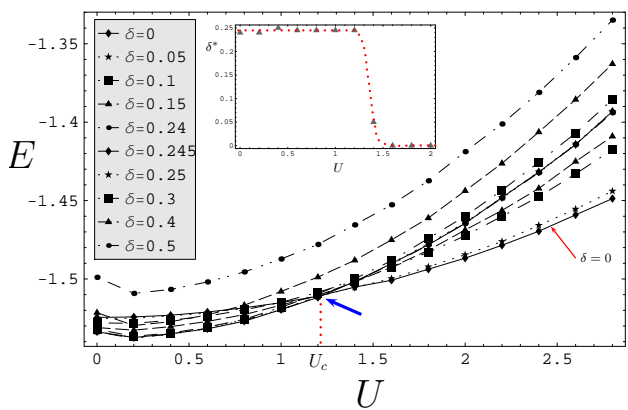}\\
  \caption{\label{fig:EvsU} Energy per site vs $U$ for several values of $\delta$
  corresponding to $\lambda=0.5$ and $\rho=0.995$. For $U_{c}\sim 1.21$ there is a
  crossing between the curves and for $U>U_{c}$ the minimum energy corresponds to $\delta=0$.
  In the inset we show the optimal value $\delta^{*}$ that minimizes the energy as a function
  of $U$. For $U>U_{c}$ the distortions are suppressed and for $0\leq U < U_{c}$ the parameter  $\delta$ takes
  an optimal value $\delta^{*}$ for most of the values of $U$ and there is one
  narrow region near $U_{c}$ where $\delta$ goes quickly to zero.    }
\end{figure}
In Figure \ref{fig:EvsU} we show the energy as a function of $U$ for
$\lambda=0.5$ and several values of $\delta$. We can see that for
$U<U_{c}$ the curves have a minimum for a non trivial value of
$\delta$ and that for $U>U_{c}$ the minimum energy is reached for
$\delta=0$. The $\delta^{*}$-value for minimum energy is plotted as
a function of $U$ in the Inset of Figure \ref{fig:EvsU} where we can
see clearly that $\delta^{*}$ stays almost constant for values of
$U$ smaller than $U_{c}$ and it tends quickly to zero in a narrow
region $U\sim U_{c}$. The critical value $U_{c}$ that we found in
our approach is lower than the one found in \cite{yuan} by means of
a MF approach, but the corresponding values of $\delta^{*}$ are very
similar.

\begin{figure}[t]
  \includegraphics[width=0.48\textwidth]{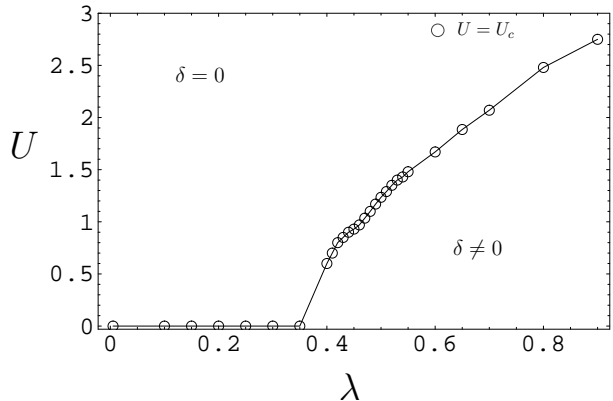}\\
  \caption{\label{fig:phases} Schematic phase diagram. The curve corresponds to $U=U_{c}$.
    }
\end{figure}

In Figure \ref{fig:phases} we show schematically the regions in the
$(\lambda,U)$ space where $\delta$ takes non trivial values, this
region is labelled as $\delta \neq 0$. The curve in Figure
\ref{fig:phases} shows the values found for $U_{c}$. For each value
of $\lambda$ we have $\delta=0$ for the $U$-values above the
critical curve. The two dimensional Hubbard model with $U>0$ at half
filling has AFM long range order, however, to depart slightly from
this point we can not ensure that this order is maintained in the
system without properly studying the AFM order parameter, but as the
main subject of this work is the study of the stability of elastic
deformations and their influence in the spectral properties, we
labeled this phase simply as $\delta=0$ or non-elastic phase. In the
region below the critical curve, where the elastic deformations
becomes stable,  the phase is labeled by $\delta\neq0$ or elastic
phase. For points close and below the critical curve there is a
small region where the $\delta$-values rise from $0$ to
$\delta^{*}$, as can be seen in the inset of Figure \ref{fig:EvsU}.


For fixed values of the Coulomb interaction the electron-lattice
constant $\lambda$ governs the phase of the system, and for this
reason it could be interesting to compare the properties of the
system at fixed $U$ for $\delta=0$ and $\delta=\delta^{*}$. In the
following we show some of the spectral properties calculated for the
model in the elastic phase.

\subsection{Spectral properties in the presence of elastic distortions .}

\begin{figure}[t]
  \includegraphics[width=0.5\textwidth]{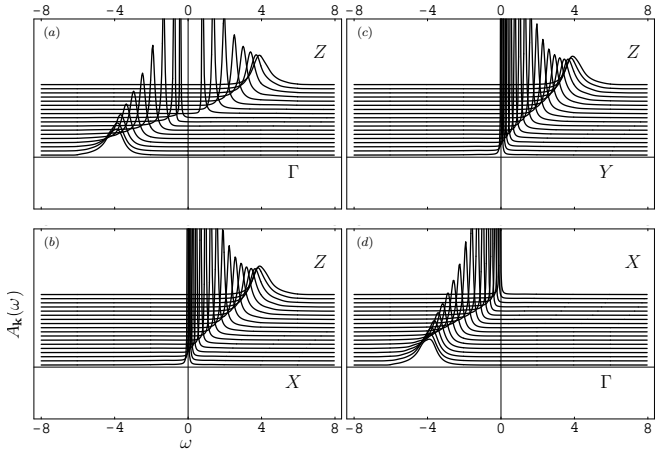}\\
  \caption{\label{fig:spectroscopy1} Spectral function for the lines $\Gamma \rightarrow Z$, $Y \rightarrow Z$, $X \rightarrow Z$
and $\Gamma \rightarrow X$. The line in Fig.\ $(a)$ is crossing
the $M$ point where there is no spectral peak as a result of the
presence of a gap in the quasiparticle dispersion. }
\end{figure}

The single-particle spectral function
$A_{\textbf{k}}(\omega)=-\frac{1}{\pi}\Im m G(\textbf{k},\omega)$ is
calculated in the second order approximation.  In Figure
\ref{fig:spectroscopy1} we show the spectral function along the
lines $\Gamma \rightarrow Z$, $Y \rightarrow Z$, $X \rightarrow Z$
and $\Gamma \rightarrow X$ for $U=1$ in the $\delta\neq0$ phase. The
behavior of the spectral function is seen to be different from the
one found in the region with $\delta=0$ \cite{zlatic}. The peaks in
the spectral function for this case are broader and asymmetric for
$\textbf{k}$ far from the FS, while for the case with distortions
the symmetry is preserved and the peaks are narrower. In the
$\delta=0$ case at higher energies the quasiparticle peaks decay
very slowly while in the $\delta\neq0$ case the peaks decay more
quickly. The splitting in the spectral function seen for $\delta=0$
is absent in the $\delta\neq0$ case. In the presence of distortions
we can see the absence of a spectral peak for some points around
$\omega=0$ which shows the existence of a gap at these points in the
renormalized dispersion. There are no possible low-energy
excitations with $\Xi\approx0$ at these points of the BZ.

\begin{figure}[t]
  \includegraphics[width=0.46\textwidth]{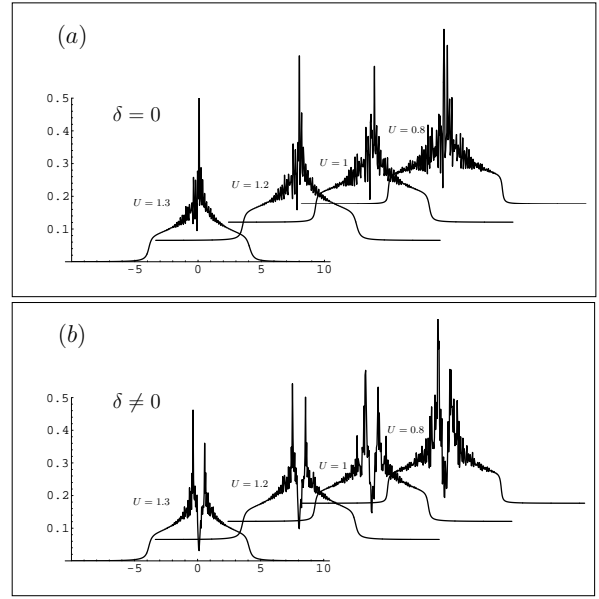}\\
  \caption{\label{fig:rho}a) Density of states $\rho(\omega)$ in the
 case with $\delta=0$  for $U=0.8$, $U=1$, $U=1.2$ and $U=1.3$. The interaction gives a transfer
  of spectral weight from low to high energies. b) Density of states $\rho(\omega)$ in the
  presence of elastic distortions  for $U=0.8$, $U=1$, $U=1.2$ and $U=1.3$.  }
\end{figure}

The density of states was calculated for fixed values of the
chemical potential $\mu$ and the Coulomb interaction $U$.
\begin{eqnarray}
g(\omega)=\frac{1}{L}\sum_{\textbf{k}}A_{\textbf{k}}(\omega)
\end{eqnarray}
In Figure \ref{fig:rho} the renormalized density of states is shown
for both phases at $\rho=0.995$ with $U=0.8$, $U=1$, $U=1.2$ and
$U=1.3$. The interaction gives a transfer of spectral weight from
low to high energies in both cases.

In the absence of distortions the weight of the logarithmic
singularity which characterizes the free system is reduced. Similar
characteristics of $g(\varepsilon)$ are obtained for the
infinite-dimensional Hubbard model \cite{bulut}. Large $U$ values
give rise to the typical two-peaks situation corresponding to the
infinite $U$ limit.

\subsection{Fermi Surface in the presence of distortions}

The solutions of $\omega-\xi_{\textbf{k}}-\Re e
\Sigma(\textbf{k},\omega)=0$ determine the renormalized dispersion
$\Xi_{\textbf{k}}$, the points where $\Xi_{\textbf{k}}=0$ define the
interacting FS. When we analyze the Hubbard Model without elastic
distortions a point with $\Xi_{\textbf{k}}<0$ is inside of the Fermi
Area and if $\Xi_{\textbf{k}}>0$ it is outside. In the present case
we need to be careful with this interpretation; a point with
$\Xi_{\textbf{k}}<0$ can be inside the Fermi Area or at the FS
depending on whether the point coincides with the gap in the
renormalized dispersion relation.

\begin{figure}[t]
  \includegraphics[width=0.48\textwidth]{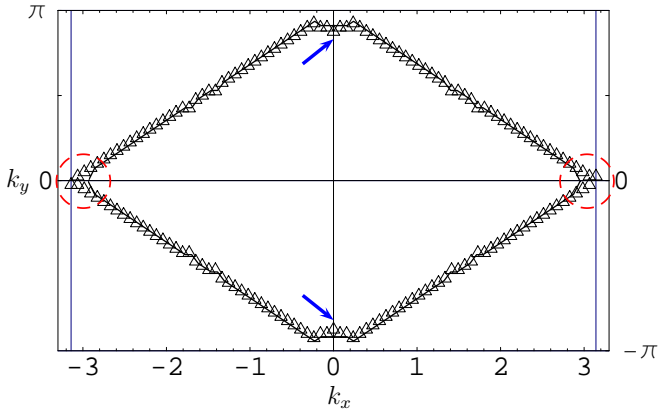}\\
  \caption{\label{fig:fs1}Interacting  Fermi Surface for $U=0.8$, $\mu=-0.1$ and $\delta=0.24$.
   The red circles mark the zones where the surface stretches in the $k_{x}$-direction
   and the blue arrows the zones where the surface is contracted in the $y$-direction.  }
\end{figure}

The interacting FS for $\delta=0.24$ and $\mu=-0.1$ is shown in
Figure \ref{fig:fs1}. The symbols correspond to $U=0.8$ and the
solid line to $U=0$. We can see that the effect of the interaction
leads to an anisotropic surface resembling a nematic phase FS. The
points in the interacting FS evolve so that the point on the
$k_{x}$-axis comes closer to the $X$ point and the point on the
$k_{y}$-axis moves away from the $Y$ point when the interaction is
increased. We do not see any change in FS topology. This result is
consistent with earlier works \cite{metzner} and shows that
interactions do not modify the FS topology within the perturbatively
controlled weak coupling regime.

\section{Discussion and Summary}

The  Hubbard model on the square lattice in the weak coupling regime
in the presence of lattice distortions that follow Peierls-like
patterns was investigated.

Using second order perturbation theory, several spectral properties
are calculated and compared with the Hubbard model in the absence of
lattice distortions. The stability of the distortions was analyzed
as a function of $U$ and $\lambda$ finding that the Coulomb
interaction suppresses the lattice distortions for values of $U_{c}$
smaller than previous MF results and a schematic phase diagram is
presented.

The results show that the Interacting Fermi Surface is anisotropic
in the presence of distortions, even in the weak coupling regime.
The Fermi Surface topology  does not change in any of the two
phases. This result is consistent and complementary with earlier
results \cite{metzner} where  the stability of the FS topology in
the absence of distortion has been analyzed previously by Metzner
{\it et al} in the weak coupling regime . The results presented in
this paper show a similar behavior in the presence of distortions.

\section*{ACKNOWLEDGMENTS}
We would like to thank D.C.\ Cabra, G. L.  Rossini and H.D.\ Rosales
for helpful discussions. This work was partially supported by
ECOS-Sud Argentina-France colaboration (Grant No A04E03), PICS
CNRS-Conicet (Grant No. 18294), PICT ANCYPT (Grant No 20350), and
PIP CONICET (Grant No. 5037).

%
%
%


\begin{thebibliography}{}
\bibitem{nota2} The pattern (a) has two possible accomplishments, with $\delta^{x}=0$ or $\delta^{y}=0$.
\end{thebibliography}
\end{document}